\documentclass[10pt,conference]{IEEEtran}
\makeatletter
\@ifpackageloaded{hyperref}{
  \hypersetup{draft}
}{
  \PassOptionsToPackage{draft}{hyperref}
}
\makeatother

\usepackage{xcolor,soul,framed} 
\usepackage[noadjust]{cite}
\usepackage[
bookmarks=false,
hidelinks
]{hyperref}
 
\colorlet{shadecolor}{yellow}

\usepackage[pdftex]{graphicx}
\graphicspath{{../pdf/}{../jpeg/}}
\DeclareGraphicsExtensions{.pdf,.jpeg,.png}
\usepackage[caption=false, font=footnotesize]{subfig}
\usepackage{amssymb,mathtools}
\usepackage{array}
\usepackage{mdwmath}
\usepackage{mdwtab}
\usepackage{eqparbox}
\usepackage{url}

\usepackage[ruled,vlined,linesnumbered]{algorithm2e}
\usepackage{enumitem}
\usepackage[mathcal]{euscript} 
\usepackage{tabularx} 
\usepackage{multirow} 
\usepackage{booktabs} 
\usepackage{makecell} 
\usepackage{aligned-overset} 
\usepackage{multicol}

\usepackage{balance}
\usepackage{bbm}
\usepackage{amsmath}
\usepackage{amsthm}
\usepackage{pifont}

\graphicspath{{./Figures/}}

\renewcommand{\vec}[1]{\boldsymbol{\mathrm{#1}}}

\newcommand{\RIS}{\mathtt{RIS}}

\newcommand{\user}{\mathtt{IU}}

\newcommand{\ap}{\mathtt{AP}}

\newcommand{\pl}{\ell}

\usepackage[hidelinks]{hyperref}
\usepackage{xcolor}

\begin{document}
      \title{RIS-Assisted Joint Resource Allocation for 6G FR3 IoT Networks}
\author{
  \IEEEauthorblockN{Muddasir Rahim, Irfan Azam, and Soumaya Cherkaoui}
  \IEEEauthorblockA{Department of Computer Engineering and Software Engineering, Polytechnique Montreal, Canada} \IEEEauthorblockA{Emails: muddasir.rahim@polymtl.ca, irfan.azam@polymtl.ca, soumaya.cherkaoui@polymtl.ca}
}

\maketitle
\begin{abstract}
In sixth-generation (6G) networks, the deployment of large numbers of Internet of Things (IoT) users (IU) necessitates efficient resource utilization and reliable connectivity, making resource allocation a critical factor. Specifically, the upper mid-band (FR3) spectrum has emerged as a promising candidate for 6G systems due to its favorable balance between bandwidth availability and coverage. However, translating these spectral advantages into performance gains in dense IoT environments requires intelligent management of interference and propagation impairments. In this paper, we propose a reconfigurable intelligent surface (RIS)-assisted IoT network operating in the FR3 band to enhance coverage and improve signal quality. Furthermore, we formulate a joint power allocation and IU-RIS association problem to maximize the achievable sum rate under practical channel conditions and power constraints. The resulting problem is nonconvex and combinatorial due to interference coupling and binary association variables. To address this challenge, we develop a multiphase resource allocation framework that integrates a successive convex approximation (SCA)-based power allocation scheme combined with a matching-theory-based user association algorithm. Simulation results demonstrate that the proposed scheme significantly outperforms conventional greedy and random search schemes in terms of sum-rate enhancement.
\end{abstract}
\section{INTRODUCTION}\label{introduction}
\IEEEPARstart{T}{he} rapid expansion of Internet of Things (IoT) applications has introduced diverse and demanding requirements for wireless communication networks, including ultra-high data rates, low latency, and high reliability~\cite{khan2022urllc,ranjha2022urllc,rahim2026reliable}. Sixth-generation (6G) networks must have greater capacity to accommodate these expanding demands and new data intensive applications, such as immersive extended/mixed reality (XR/MR) and ultra-high-definition video streaming. Moreover, 6G must remain commercially viable while accommodating a wide range of new use cases and functionalities, ensuring reliable connectivity for a massive number of heterogeneous users \cite{pala2023spectral}. Using higher frequency bands with wider bandwidths, which support higher data rates, is one way to meet these needs. Therefore, there is growing interest from both academia and industry in identifying suitable frequency bands, with particular attention to the upper mid-band (UMB) (7–15 GHz), also known as frequency range 3 (FR3)~\cite{chaves2024coverage}. Moreover, the FR3 band is now referred to as the "Golden Band" for 6G, owing to its promising balance between bandwidth availability and coverage~\cite{cui20236g,rahim2026dual}. However, translating the spectral advantages of FR3 into reliable large-scale coverage requires overcoming non-negligible propagation loss and frequent line-of-sight (LoS) blockages in cluttered environments, motivating the need for environment shaping technologies beyond dense small-cell deployments. 

To overcome propagation loss and enhance coverage, one proposed solution is the use of reconfigurable intelligent surfaces (RISs). RISs, particularly passive forms, provide a low-cost, energy-efficient alternative by enabling reconfigurable radio environments through large arrays of passive reflecting elements~\cite{rahim2024joint,rahim2024multi,rahim2024user,huang2019reconfigurable,rahim2024jpusa,tashman2025dynamic,tashman2026trustworthy}. Despite growing interest in FR3 communications for 6G systems, most existing work focuses on enhancements to the multiple-input-multiple-output (MIMO) architecture, beam management strategies, or coverage evaluations. The integration of RISs in the FR3 band remains largely underexplored, particularly in dense IoT deployments where large numbers of low-power devices coexist and compete for limited radio resources. Moreover, most RIS-related studies focus on the FR2 or THz bands, leaving the resource management challenges in FR3-based IoT networks uninvestigated.

\subsection{Related Works}
In recent works, FR3 has emerged as an attractive spectrum option for 6G because it offers a practical compromise, such as a wider bandwidth than conventional sub-6 GHz bands, while avoiding some of the severe propagation issues seen at much higher frequencies. This has motivated growing interest in understanding FR3 in terms of achievable performance and possible application scenarios. However, dedicated studies on MIMO operation in FR3 remain relatively limited~\cite{heath2024beamsharing,bjornson2025enabling,tian2024mid}. For example,~\cite{heath2024beamsharing} considered a MIMO system serving both near-field and far-field users, and proposed a user pairing strategy together with beam sharing to support such heterogeneous users. In~\cite{bjornson2025enabling}, the authors introduced the notion of gigantic MIMO (gMIMO), referring to arrays with at least 256 antenna ports, to emphasize deployments that go beyond typical ultra-massive MIMO configurations. This work then examined the opportunities and feasibility of gMIMO specifically in the FR3 band. In addition,~\cite{tian2024mid} investigated the use of extra-large-scale MIMO (XL-MIMO) for mid-band communications, focusing on developing an analytical model to evaluate key network-level metrics such as spectral efficiency and outage probability.

Regarding RIS-related literature,~\cite{mohsan2023irs} provides a detailed survey on RIS-assisted UAV communications, covering system architectures, channel modeling aspects, and common design goals. The authors of~\cite{abdalla2020uavs} further highlighted the potential of combining RIS with UAV platforms in future cellular networks, discussing representative use cases, major challenges, and open research problems, along with related studies on spectrum sharing, physical-layer security, access enhancement, and coverage extension. A more design-oriented contribution is presented in~\cite{bui2025joint}, where a joint optimization framework was proposed for transmit power allocation and RIS phase-shift tuning, aiming to improve energy efficiency while ensuring reliable service for all users. In this setup, RIS panels placed on buildings help extend coverage by creating an additional reflected path between UAVs and ground users. Meanwhile,~\cite{farre2025dynamic} examined the integration of RISs in combined non-terrestrial and terrestrial networking, targeting improved connectivity and capacity in dense environments by optimizing coverage as well as interference suppression. The work in~\cite{you2022enabling} also introduced two complementary schemes: RIS-assisted UAV links and aerial-RIS-assisted terrestrial links, illustrating how UAVs and RISs can be jointly exploited to support integrated aerial–terrestrial communications in next-generation networks.

Despite these advances, RIS research in FR3 is still in its early stages. To the best of the authors’ knowledge,~\cite{kara2024reconfigurable} is one of the few works that explicitly demonstrates the relevance of RISs in the FR3 band. It investigates the conditions under which RISdeployment can yield noticeable gains and discusses effective deployment strategies to enhance FR3 system performance. Overall, the RIS literature remains heavily concentrated on FR2 (mmWave/THz) and primarily single-tier terrestrial scenarios, while multi-tier operation in FR3, including the combined use of aerial and terrestrial IRSs, has received comparatively little attention, even though FR3 is becoming increasingly important for 6G and exhibits different coverage and blockage trade-offs.

\subsection{Contributions}\label{contributions}
Existing resource allocation approaches in RIS-assisted networks often address power control or user association separately. However, in realistic IoT environments, transmit power allocation and RIS association are inherently coupled due to interference interactions and RIS-dependent effective channel gains. To the best of our knowledge, this is the first work to integrate an SCA combined with matching-theory-based resource allocation in RIS-assisted IoT networks in UMB spectrums.  Given this, the contributions of this work are as follows:

\begin{itemize}
  \item We propose a practical system model for RIS-assisted IoT networks operating in the UMB spectrum, in which IoT devices communicate in the downlink.
  \item We formulate the resource allocation problem with the objective of maximizing the achievable sum rate. To solve this problem, we develop an SCA-based power allocation scheme combined with a matching-theory-based user association algorithm under a realistic channel model.
  \item Simulation results demonstrate that the proposed scheme achieves performance close to the ES scheme, while significantly outperforming GS-based and random search schemes.
\end{itemize}
\section{System Model} \label{model}
\subsection{Network Model}
We consider a network consisting of a single AP equipped with $N$ antennas serving a $K$ set of IoT users (IUs). IUs transmit their signals either through direct links to the AP or via one of the available RISs when the direct path is weak or blocked. The channels between the AP and the RISs, as well as between the RISs and the IUs, are modeled as quasi-static fading channels in the FR3 band. To ensure reliable communication in environments where direct links may be obstructed, the network is supported by $L$ RISs. The RIS set is denoted by $\mathcal{L} = \{\RIS_1, \RIS_2, \ldots, \RIS_l, \ldots, \RIS_L\}$. Each RIS comprises $M = M_y M_z$ passive reflectors arranged in a uniform planar array, where $M_y$ and $M_z$ denote the number of reflectors along the horizontal and vertical directions, respectively. The center locations of the AP, $k^{th}$ IU ($\user_{k}$), and $l^{th}$ RIS ($\RIS_{l}$) are denoted by ${\mathbf p_{b} = (x_{b}, y_b, z_{b})}$, ${\mathbf p_{k} = (x_{k}, y_{k}, z_{k})}$, and ${\mathbf p_{l} = (x_{l}, y_{l}, z_{l})}$, respectively. Furthermore, the distance from the AP to $\user_{k}$ is $d_{b, k}\triangleq \Vert \mathbf{p}_{b}  - \mathbf{p}_{k}\Vert$, from the AP to the center of $\RIS_{l}$ is $d_{b, l}\triangleq \Vert \mathbf{p}_{b}  - \mathbf{p}_{l} \Vert$, and from $\RIS_{l}$ to $\user_{k}$ is denoted as $d_{l, k}\triangleq \Vert \mathbf{p}_{l}  - \mathbf{p}_{k} \Vert$.
\begin{figure}[!t]
     \centering
\includegraphics[width=\linewidth]{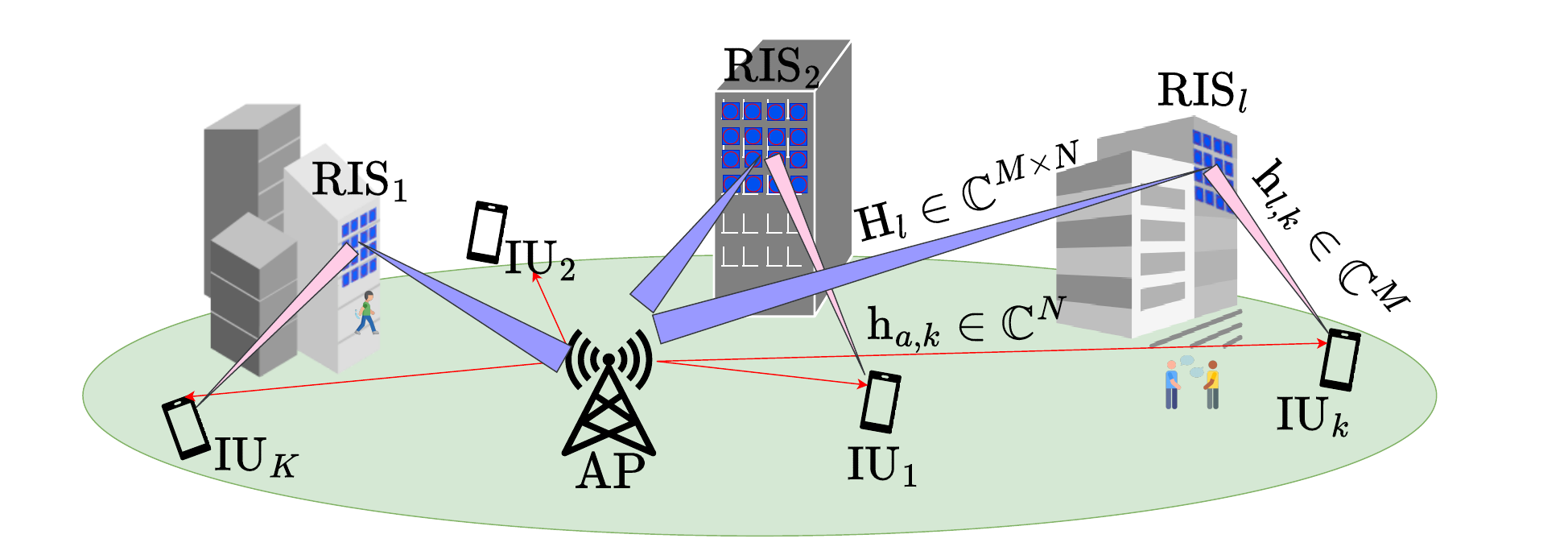}
    \caption{System model of RIS-assisted IoT network.
     }
     \label{m1}   
\end{figure}

\subsection{Channel Model}
We consider an IoT network transmitting data in downlink communication with both the direct link and the reflecting links from RISs. Let $\vec{h}_{a,k}\in \mathbb{C}^{N}$ is the direct channel vector from $\ap_n$-to-$\user_k$. which can be written as
\begin{align}\label{channel_user-bs}
   [\vec{h}_{a,k}]_{n} =h_{a,n,k} = \sqrt{\pl_{k,a,n}} \, e^{-j \omega d_{a,k}},
 \end{align}
where $\pl_{k,a,n}$ is the pathloss from $\user_{k}$-to-$\ap_n$ and $\omega = \frac{2\pi f}{c}$ is the wave number at frequency $f$. The channel from $n^{th}$ antenna of AP ($\ap_n$)-to-$\RIS_{l,m}$ can be expressed as
 \begin{align}\label{channel_bs-irs}
   [\vec{H}_{l}]_{m,n} =h_{l,m,n} = \sqrt{\pl_{l,m,n}} \, e^{-j \omega d_{b,l}},
 \end{align}
where $\pl_{l,m,n}$ is the pathloss from $\RIS_{l,m}$-to-$\ap_n$ and $\vec{H}_{l}\in \mathbb{C}^{M \times N}$ is the channel matrix from $\ap$-to-$\RIS_l$. Finally, $\vec{h}_{l,k} \in \mathbb{C}^{M}$ denotes the channel vector from $m^{th}$ element of $\RIS_l$ ($\RIS_{l,m}$)-to-$\user_k$, which can be expressed as
\begin{align}\label{channel_irs_ue}
   [\vec{h}_{l,k}]_m= h_{l,m,k} = \sqrt{\pl_{l,m,k}} \, e^{-j \omega d_{l,k}},
\end{align}
where $\pl_{l,m,k}$ is the pathloss from $\RIS_{l,m}$-to-$\user_k$. 

 Let $\vec{\Gamma}$ be the association matrix and $\gamma_{k,l}\in\{0,1\}$ be a binary association variable, which can be written as

\begin{align}
    \gamma_{k,l} =
\begin{cases}
1, & \user_k \text{ is associated with } \RIS_l, \\
0, & \user_k \text{ otherwise}.
\end{cases}
\end{align}

Then, the downlink effective channel vector from $\ap$-to-$\user_k$ is  $\vec{h}_{k}\in \mathbb{C}^{N}$, which can be written as
\begin{align}\label{eq_channel}
    \vec{h}_{k} =\underbrace{\vec{h}_{a,k}}_{\text{direct channel vector}} +\underbrace{\sum_{l=1}^{L}\gamma_{k,l} \vec{H}_{b,l}^{\sf H} \, \vec{\Theta}_l \, \vec{h}_{l,k}}_{\text{cascaded channel vector}},
 \end{align}
 where $\vec{\Theta}_l$ is the RIS reflection matrix, which can be written as 
\begin{equation}\label{reflection_matrix}
    \vec{\Theta}_l = \mathrm{diag}([\kappa_{l,1} e^{j \theta_{l,1}}, \ldots,\kappa_{l,m} e^{j \theta_{l,m}}, \ldots, \kappa_{l,M} e^{j \theta_{l,M}}]),
\end{equation}
where $\kappa_{l,m} \in [0,1]$ and $\theta_{l,m} \in [0, 2\pi)$ denote amplitude coefficients and phase shifts. 
\subsection{Signal Model}
We consider a precoding matrix ${\vec{W}} \in \mathbb{C}^{N \times K}$ with columns ${\vec{w}}_k \in \mathbb{C}^{N}$ denoting unit-norm beam direction for $\user_k$, which can be expressed as ${\vec{w}}_k = \sqrt{p_k}\widehat{\vec{w}}_k,$ where $p_k$ and $\widehat{\vec{w}}_k$ are the transmit power scaling factor and the beamforming vector for $\user_k$, respectively. For the given power budget $\vec{P}_{\max}$, the power constraint can be expressed as
\begin{align}
  \textstyle \sum_{k=1}^K \Vert{\vec{w}}_k\Vert^2 = \mathsf{Tr} \{ \vec{\widehat{W}}\vec{P} \vec{\widehat{W}}^{\sf H}\} \leq \vec{P}_{max},
\end{align}
where $\vec{P} = \mathrm{diag}(p_1,\ldots,p_k,\ldots,p_K)$.
Let $\vec{x} \in \mathbb{C}^{N}$ be the transmit signal vector from the AP to IUs, where
$s_k$ is the data symbol intended for $\user_k$, then, the transmitted signal vector from the AP can be written as 
\begin{align} \label{transmit_signal}
\textstyle\vec{x}=\sum_{{k=1 }}^K{\vec{w}}_k s_k=\sum_{{k=1  }}^K\sqrt{p_k}\widehat{\vec{w}}_k s_k.
\end{align}
The received signals at $\user_k$ via $\RIS_l$ can be expressed as 
\begin{align}\label{reciver_1}
  \textstyle   y_{l,k} &= \sqrt{p_{k}}(\mathbf{h}_{k})^H \widehat{\vec{w}}_k s_k + \sum_{i \neq k}^{K}\sqrt{p_{i}}(\mathbf{h}_{k})^H \widehat{\vec{w}}_i s_i + n_k\nonumber\\ &=\underbrace{\sqrt{p_{k}}(\vec{h}_{a,k} +\sum_{l=1}^{L}\gamma_{k,l} \vec{H}_{b,l}^{\sf H}  \vec{\Theta}_l \vec{h}_{l,k})^H \widehat{\vec{w}}_k s_k}_{\text{desired signal}}\nonumber\\ & + \underbrace{\sum_{i \neq k}^{K}\sqrt{p_{i}}(\vec{h}_{a,k} +\sum_{l=1}^{L}\gamma_{i,l} \vec{H}_{b,l}^{\sf H}  \vec{\Theta}_l \vec{h}_{l,k})^H \widehat{\vec{w}}_i s_i}_{\text{interference}} + \underbrace{n_k}_{\text{noise}},
\end{align}

  Based on the received signal in~(\ref{reciver_1}), we derive the signal-to-noise-plus-interference ratio (SINR) and the data rate of $\user_k$, respectively as
   \begin{align}\label{sinr}
     \Lambda_k=\frac{p_k\Vert(\vec{h}_{a,k} +\sum\limits_{\substack{l=1}}^L\gamma_{k,l} \vec{H}_{b,l}^{\sf H}  \vec{\Theta}_l \vec{h}_{l,k})^H \widehat{\vec{w}}_k \Vert^2 }{\sum\limits_{\substack{i\neq k}}^K\! p_i\big(\Vert(\vec{h}_{a,k} +\sum\limits_{\substack{l=1}}^L\gamma_{i,l} \vec{H}_{b,l}^{\sf H}  \vec{\Theta}_l \vec{h}_{l,k})^H \widehat{\vec{w}}_i\Vert^2\big)\!+\!\sigma_k^2},
 \end{align}
  \begin{align}\label{data_rate}
     &R_k=\log_2(1 +\Lambda_k)\nonumber\\&=\log_2\!\!\Bigg(\!1\!\!+\!\frac{p_k\Vert(\vec{h}_{a,k} +\sum\limits_{\substack{l=1}}^L\gamma_{k,l} \vec{H}_{b,l}^{\sf H}  \vec{\Theta}_l \vec{h}_{l,k})^H \widehat{\vec{w}}_k \Vert^2 }{\sum\limits_{\substack{i\neq k}}^K\! p_i\big(\Vert(\vec{h}_{a,k} \!+\!\sum\limits_{\substack{l=1}}^L\gamma_{i,l} \vec{H}_{b,l}^{\sf H}  \vec{\Theta}_l \vec{h}_{l,k})^H \widehat{\vec{w}}_i\Vert^2\big)\!+\!\sigma_k^2}\!\Bigg).
 \end{align}
 
\subsection{Optimization Problem Formulation}\label{pform}
In massive IoT networks, multiple resource-management challenges arise, including interference and network topology. In this work, we will maximize the network achievable sum rate by optimizing the transmit power and user IU-RIS association. Then, the optimization problem can be established to maximize the network sum rate as 
\begin{subequations}\label{eq_opt_prob}
\begin{alignat}{2}
& \underset{\vec{P, \,\Gamma}}{\text{ maximize}}
&
&\quad\sum_{k=1}^K R_k (\vec{P, \Gamma}), \label{eq_optProb}\\ 
&\text{subject to} 
 &&\quad  p_k \geq 0, \quad\forall k\in K, \label{eq_constraint1}\\
 &&&\quad 
\sum_{k=1}^K p_k\leq P_{max},\label{eq_constraint2}\\
 &&&\quad  \gamma_{k,l}\in\{0,1\}, \ \forall k,l, \label{eq_constraint3}\\
 &&&\quad  \sum_{l=1}^{L} \gamma_{k,l} \le 1,\ \forall k, \label{eq_constraint4}\\ 
 &&&\quad \sum_{k=1}^{K} \gamma_{k,l} \le 1,\ \forall l, \label{eq_constraint5}
\end{alignat}
\end{subequations}
where $P_{max}$ is the maximum transmit power budget, constraints~(\ref{eq_constraint1}) and~(\ref{eq_constraint2}) ensure that the transmit power of all users is always positive and the total power does not exceed the power budget $P_{max}$, respectively. Furthermore, constraint~\eqref{eq_constraint3}  enforces binary assignment to complete the one-to-one matching, constraint~\eqref{eq_constraint4} restricts each RIS to a single IU, and constraint~\eqref{eq_constraint5} assigns each IU to exactly one RIS.
\section{Proposed Resource Allocation Framework}\label{proposed} 
Problem~\eqref{eq_opt_prob} is NP-hard due to the presence of binary RIS association variables and the interference-coupled nonconvex rate expression. Specifically, even if the transmit powers $\vec{P}$ are fixed, the optimization over $\vec{\Gamma}$ reduces to a combinatorial association problem under constraints~\eqref{eq_constraint4} and~\eqref{eq_constraint5}, where selecting an RIS for each IU impacts the effective channels and hence the resulting rates. This structure is closely related to known NP-hard link selection problems when the objective depends on coupled utilities. Therefore, in this framework, the AP first determines the downlink transmit power and then the IU-RIS association in order to enhance the received signal quality and improve overall network performance in the FR3 band.
\begin{figure}[!t]
     \centering
\includegraphics[width=\linewidth]{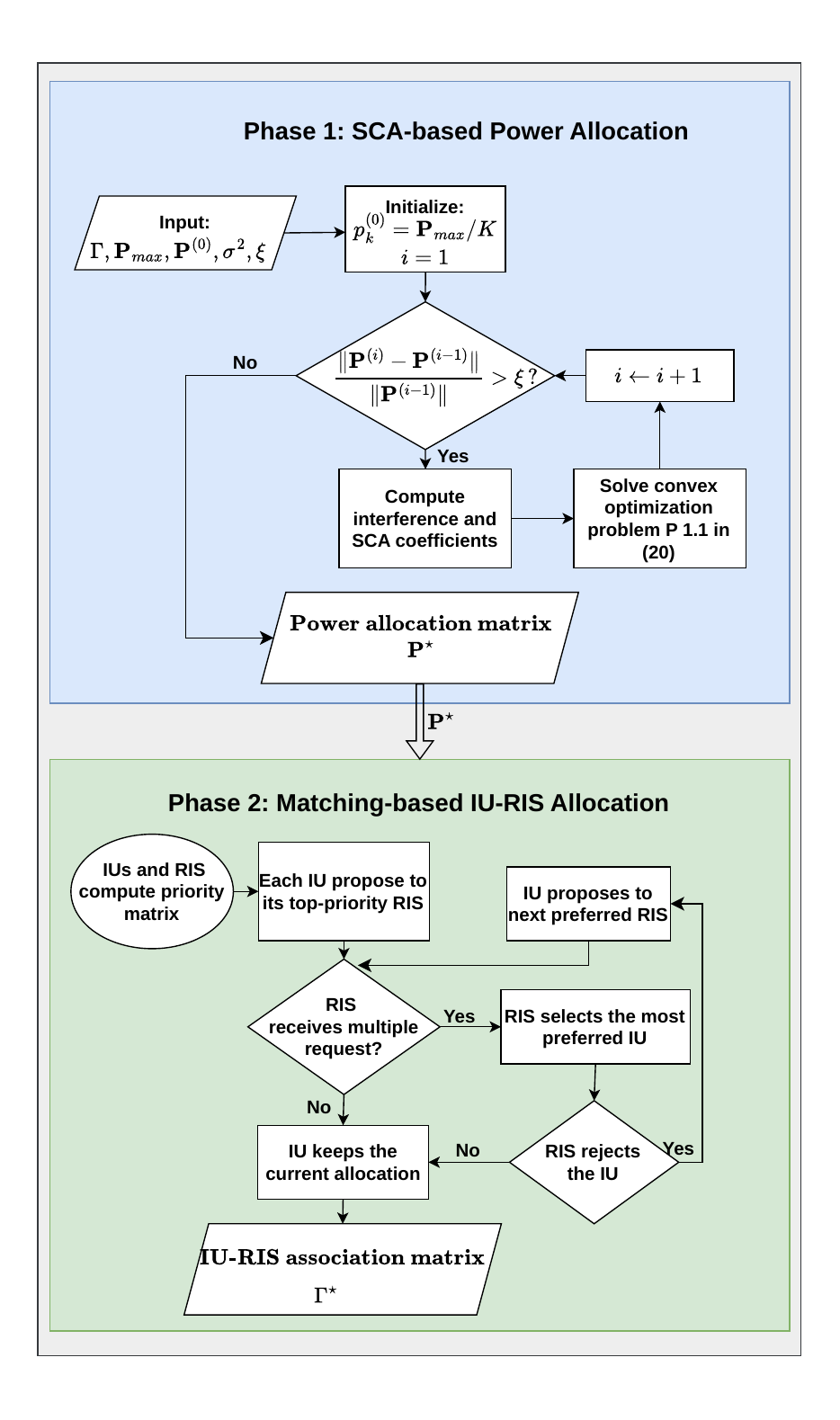}
    \caption{Proposed two-phase resource allocation algorithm: Phase 1 performs SCA-based downlink power optimization, while Phase 2 executes matching-based one-to-one IU–RIS association.}
     \label{algo}   
\end{figure}
\subsection{SCA-based Power Optimization}
Given the RIS association matrix $\vec{\Gamma}$, we optimize the downlink transmit power allocation of IUs. The power optimization subproblem is given by
\begin{subequations}\label{eq_opt_prob1.2}
\begin{alignat}{2}
& \textbf{P1 :} \underset{\vec{\mathbf{P}}}{\text{ maximize}}
&
&\quad\quad\sum_{k=1}^K R_k (\vec{ P, \Gamma}), \label{eq_optProb1.2}\\
&\text{subject to} 
 &&\quad\quad  p_k \geq 0, \quad\forall k\in K, \label{eq_constraint1.2}\\
 &&&\quad\quad 
\textstyle\sum_{k=1}^K p_k\leq P_{max},\label{eq_constraint2.2}
\end{alignat}
\end{subequations}
The objective function in~\eqref{eq_opt_prob1.2} is nonconvex due to the interference-coupled SINR, even for the given $\Gamma$. Therefore, to address this issue, we use a successive convex approximation (SCA) to solve the subproblem. The data rate formulation~\eqref{data_rate} for power allocation can be written as in~\eqref{SCA_1} on the next page.

  \begin{align}\label{SCA_1}
     &R_k(\vec{P}) =\log_2\Bigg(1 +\frac{p_k {g_{k,k}}}{\underbrace{\textstyle\sum_{ i\neq k}^K p_ig_{k,i}+\sigma_k^2}_{I_{k}(\vec{P})}}\Bigg)\nonumber\\&=\log_2\Bigg(1 +\frac{p_k {g_{k,k}}}{I_k(\vec{P})}\Bigg)=\log_2\Bigg(\frac{I_k(\vec{P})+p_k {g_{k,k}}}{I_k(\vec{P})}\Bigg),
 \end{align}  

where $g_{k,k}=\Vert(\vec{h}_{a,k} +\sum_{l=1}^{L}\gamma_{k,l} \vec{H}_{b,l}^{\sf H}  \vec{\Theta}_l \vec{h}_{l,k})^H \widehat{\vec{w}}_k  \Vert^2$ and $g_{k,i}=\Vert(\vec{h}_{a,k} +\sum_{l=1}^{L}\gamma_{i,l} \vec{H}_{b,l}^{\sf H}  \vec{\Theta}_l \vec{h}_{l,k})^H \widehat{\vec{w}}_i\Vert^2$. Now $R_k(\vec{P})$ in~\eqref{SCA_1} can be equivalently rewritten as
  \begin{align}\label{SCA_2}
R_k(\vec{P})=\log_2\Big(I_k(\vec{P})+p_k {g_{k,k}}\Big)- \log_2\Big(I_k(\vec{P})\Big).
 \end{align}

 The first term $\log_2\big(I_k(\vec{P})+p_k {g_{k,k}}\big)$ and second term $\log_2\big(I_k(\vec{P})\big)$ in \eqref{SCA_2} both are concave functions, therefore, $R_k(\vec{P})$ is difference of concave function which keeps~\eqref{SCA_2} nonconvex. Meanwhile, we utilize SCA, in which we keep the first term and linearize the second term at iteration $t$ using the first-order Taylor expansion, which can be written as
\begin{align}\label{SCA_3} \log_2\big(I_k(\vec{P})\big)\leq\log_2\big(I_k(\vec{P^{(t)}})\big)+\textstyle\sum_{i=1}^K \varrho_{k,i}^{(t)}(p_i-p_i^{(t)}),
 \end{align}
where
\begin{align}\label{SCA_4}
   \varrho_{k,i} =
\begin{cases}
\frac{1}{\ln{2}}\frac{g_{k,i}}{\log_2(I_k(\vec{P^{(t)}}))}, &  i\neq k, \\
0, &  i= k.
\end{cases}
\end{align}
Since $\log_2\big(I_k(\vec{P})\big)$ is concave and the first-order Taylor expansion in \eqref{SCA_3} provides an upper bound which guarantees monotonic improvement in SCA. Utilizing \eqref{SCA_2} and \eqref{SCA_3} we have 
\begin{align}\label{SCA_5}
R_k(\vec{P})=&\log_2\Big(I_k(\vec{P})+p_k {g_{k,k}}\Big)\nonumber\\&- \log_2\big(I_k(\vec{P^{(t)}})\big)-\sum_{i=1}^K \varrho_{k,i}^{(t)}(p_i-p_i^{(t)})\nonumber\\&=\log_2\Big(\sum\limits_{\substack{ i\neq k}}^K \alpha_j p_ig_{k,i}+\sigma_k^2+p_k {g_{k,k}}\Big)\nonumber\\&- \underbrace{\log_2\big(I_k(\vec{P^{(t)}})\big)+\sum_{i=1}^K \varrho_{k,i}^{(t)}p_i^{(t)}}_{\text{constant}}-\sum_{i=1}^K \varrho_{k,i}^{(t)}p_i.
 \end{align}
Constant terms that are independent of $\vec{P}$ are excluded since they do not affect the optimization. Thus, we can rewrite the optimization problem $\vec{P1}$ in \eqref{eq_opt_prob1.2} using \eqref{SCA_5}, which can be written as
\begin{subequations}\label{eq_opt_prob1.2.1}
\begin{alignat}{2}
& \textbf{P1.1 :} \underset{\vec{\mathbf{P}}}{\text{ maximize}}
& 
&\,\sum_{k=1}^K\Big[ \log_2\big(\sum\limits_{\substack{ i\neq k}}^K p_ig_{k,i}+\sigma_k^2+p_k {g_{k,k}}\big)\nonumber\\&\quad\quad\quad-\sum_{i=1}^K \varrho_{k,i}^{(t)}p_i\Big], \label{eq_optProb1.2.1}\\ 
&\text{subject to} 
 &&\quad\quad  p_k \geq 0, \quad\forall k\in K, \label{eq_constraint1.2.1}\\
 &&&\quad\quad 
\textstyle\sum_{k=1}^K p_k\leq P_{max},\label{eq_constraint2.2.1}
\end{alignat}
\end{subequations}

Problem $\vec{P1.1}$ is a convex optimization problem, which can be efficiently solved using standard convex solvers such as CVX.

\subsection{Matching-based IU-RIS Association}
Then, we use the power allocation matrix obtained in subproblems $\vec{P1.1}$, to formulate the IU-RIS association as
\begin{subequations}\label{eq_opt_prob1.3}
\begin{alignat}{2}
& \textbf{P2 :} \underset{\vec{\Gamma}}{\text{ maximize}}
& 
&\quad\quad\sum_{k=1}^K  R_k (\vec{P^{\star}, \Gamma}), \label{eq_optProb1.3}\\
&\text{subject to} 
 &&\quad\quad p_k \geq 0, \quad\forall k\in K, \label{eq_constraint1.3}\\
 &&&\quad\quad 
\textstyle\sum_{k=1}^K p_k\leq P_{max},\label{eq_constraint2.3}\\
 &&&\quad\quad  \gamma_{k,l}\in\{0,1\}, \ \forall k,l, \label{eq_constraint3.3}\\
 &&&\quad\quad  \textstyle\sum_{l=1}^{L} \gamma_{k,l} \le 1,\ \forall k, \label{eq_constraint4.3}\\ 
 &&& \quad\quad \textstyle\sum_{k=1}^{K} \gamma_{k,l} \le 1,\ \forall l, \label{eq_constraint5.3}
\end{alignat}
\end{subequations}
We deploy multiple RISs to assist the downlink communication. The IU-RIS association problem $\vec{P2}$ is formulated as a one-to-one matching problem, where each IU can be associated with at most one RIS within a transmission slot. As illustrated in Phase 2 of Fig.~\ref{algo}, the association procedure follows an IU-proposing deferred-acceptance mechanism. Initially, all IUs are unmatched and construct preference lists of RISs, ranked by their achievable data rates. In each iteration, every unmatched IU proposes to its most preferred RIS that has not yet rejected it. Upon receiving one or more proposals, each RIS evaluates the achievable data rates of the proposing IUs and tentatively accepts the IU with the highest rate, rejecting the remaining proposals. If a RIS is already matched, it compares the new proposal with its current match and retains the IU that yields the higher data rate. Rejected IUs remove the corresponding RIS from their preference lists and proceed to propose to their next preferred RIS in subsequent iterations. This iterative process, as depicted in Phase 2 of the flowchart, continues until no further proposals can be made, at which point a stable IU-RIS matching is obtained.

\section{Simulation Results and Discussion} \label{pref}
We consider a network with multiple users and a single multi-antenna AP deployed over the network area. Several RISs are installed, and each RIS comprises $100\times100$ reflecting elements with half-wavelength element spacing ($M_y=M_z= \lambda/2$). The simulation parameters are summarized in Table~\ref{tab:sim}. All algorithms are implemented in MATLAB. Unless otherwise stated, each plotted point is averaged over $10^6$ independent channel realizations. We evaluate the proposed method against two benchmark schemes, where the transmit power allocation in all cases is obtained using the SCA-based optimization framework. First, in the SCA-GS-based allocation, each IU selects the RIS that yields the highest achievable data rate. If a RIS receives multiple proposals, one IU is selected randomly, while the others are rejected. Second, in the SCA-RS baseline, the IU–RIS association matrix is generated randomly, whereas the transmit power is still optimized using the SCA procedure.
\begin{table}[!t]
\centering
\renewcommand{\arraystretch}{1.5}
\setlength{\tabcolsep}{10pt}
\caption{Simulation parameters}
\label{tab:sim}
\begin{tabular}{l @{\hspace{10mm}} l}
\hline
\textbf{Parameters} & \textbf{Values} \\ \hline
Carrier frequency ($f_c$) {[}GHz{]} & $15$ \\
Number of antennas at AP & $64$ \\
Number of IUs & $5$ \\
Network area ${[}\text{m}^2{]}$& $100$ \\
AP power budget {[}dBm{]} & $23$~\cite{rahim2023joint} \\
Power density of noise {[}dBm/Hz{]} & $-174$ \\
Number of reflectors & $100\times100$~\cite{rahim2025hierarchical} \\
Channel bandwidth {[}MHz{]} & $400$ \cite{karaman2025demand} \\
Noise figure {[}dB{]} & $10$ \\
Side length of reflectors {[}m{]} & ${\lambda}/{2}$ \\ \hline
\end{tabular}
\end{table}
\begin{figure}[!htp]
     \centering
    \includegraphics[width=\linewidth]{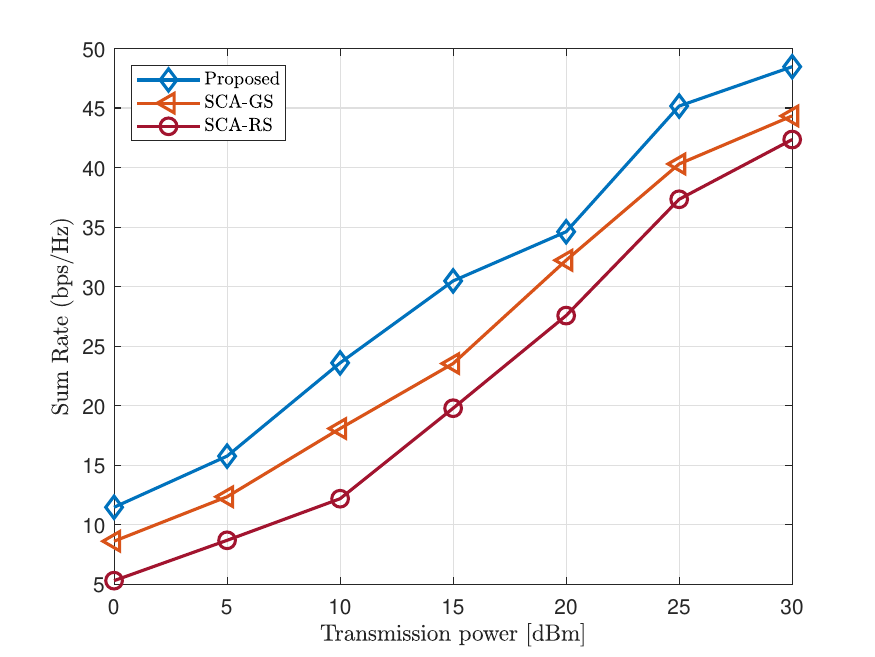}
     \caption{Achievable network sum rate as a function of the AP power budget.}
     \label{rate_power}   
\end{figure}

The sum rate as a function of the AP transmit power is shown in Fig.~\ref{rate_power}. As expected, the sum rate increases with the transmit power for all considered schemes due to the improved received signal strength and enhanced SINR conditions. Notably, the proposed matching-based scheme consistently achieves higher sum rates compared with the SCA-GS and SCA-RS baseline methods across the entire power range. This improvement mainly stems from the more efficient IU–RIS association enabled by the matching framework, which better exploits channel conditions and resource availability. Consequently, the proposed approach provides more effective interference management and resource utilization, resulting in superior overall network sum rate.

\begin{figure}[!htp]
     \centering
\includegraphics[width=\linewidth]{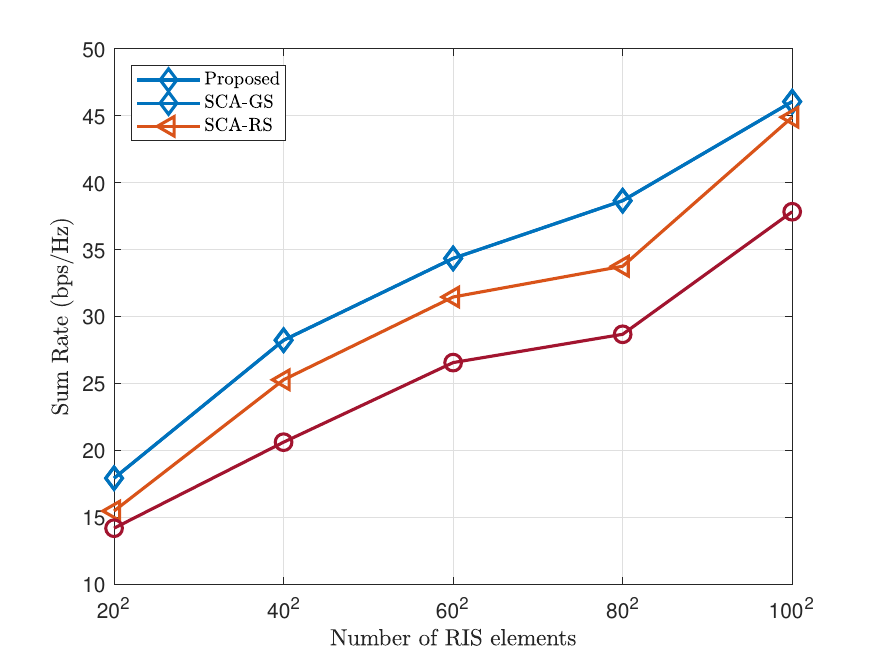}
    \caption{Achievable network sum rate as a function of the number of RIS elements.}
     \label{rate_elements}   
\end{figure}

The sum rate as a function of the number of RIS elements is illustrated in Fig.~\ref{rate_elements}. Particularly, increasing the number of RIS elements improves the sum rate for all considered schemes due to the enhanced passive beamforming gain and improved signal reflection capability. The proposed matching-based scheme consistently achieves higher performance compared with the SCA-GS and SCA-RS baseline methods across all RIS configurations. This gain is mainly attributed to the more efficient IU–RIS association enabled by the matching framework, which better aligns IUs with favorable channel conditions and available RIS resources. 

\section{Conclusions}\label{conc}
The large-scale integration of IUs in 6G networks further elevates the need for efficient resource utilization and dependable connectivity. This work investigated an RIS-assisted IoT network operating in the FR3 spectrum. Within this framework, the power allocation and IU-RIS association problem is formulated to maximize the network achievable sum rate. To address this problem, we developed an SCA-based power allocation scheme combined with a matching-theory-based IU-RIS association algorithm. Simulation results demonstrate that the proposed scheme significantly outperforms conventional greedy and random search schemes in terms of sum-rate enhancement. Future research will extend the proposed framework to incorporate active IU detection and reliability-aware resource allocation for grant-free massive IoT scenarios. Specifically,  joint optimization of active IU detection, power control, and IU-RIS association under dynamic traffic conditions represents a promising direction for enhancing scalability and robustness in RIS-assisted FR3 networks.
\bibliographystyle{IEEEtran}
\balance
\bibliography{References}

@article{karaman2025demand,
  title={{On-Demand HAPS-Assisted Communication System for Public Safety in Emergency and Disaster Response}},
  author={Karaman, Bilal and Ba{\c{s}}t{\"u}rk, Ilhan and Kara, Ferdi and Zeydan, Engin and Beyaz{\i}t, Esra Aycan and Ta{\c{s}}k{\i}n, Sezai and Bj{\"o}rnson, Emil and Yanikomeroglu, Halim},
  journal={arXiv preprint arXiv:2507.09153},
  year={2025}
}

@article{huang2019reconfigurable,
  title={{Reconfigurable intelligent surfaces for energy efficiency in wireless communication}},
  author={Huang, Chongwen and Zappone, Alessio and Alexandropoulos, George C and Debbah, M{\'e}rouane and Yuen, Chau},
  journal={IEEE Trans. Wireless Commun.},
  volume={18},
  number={8},
  pages={4157--4170},
  year={2019},
  publisher={IEEE}
}

@article{ranjha2022urllc,
  title={{URLLC} in {UAV}-enabled multicasting systems: A dual time and energy minimization problem using {UAV} speed, altitude and beamwidth},
  author={Ranjha, Ali and Kaddoum, Georges and Rahim, Muddasir and Dev, Kapal},
  journal={Computer Communications},
  volume={187},
  pages={125--133},
  year={2022},
  publisher={Elsevier}
}

@article{rahim2023joint,
  title={{Joint Devices and {IRSs} Association for Terahertz Communications in Industrial IoT Networks}},
  author={Rahim, Muddasir and Kaddoum, Georges and Do, Tri Nhu},
  journal={IEEE Trans. Green Commun. Networking	},
  year={2023},
  publisher={IEEE}
}

@article{rahim2024multi,
  title={{Multi-IRS-Aided Terahertz Networks: Channel Modelling and User Association With Imperfect CSI}},
  author={Rahim, Muddasir and Nguyen, Thanh Luan and Kaddoum, Georges and Do, Tri Nhu},
  journal={IEEE Open J. Commun. Soc.	},
  year={2024},
  publisher={IEEE}
}

@article{chaves2024coverage,
  title={{Coverage evaluation of 7--15 GHz bands from existing sites}},
  author={Chaves, F and Chizhik, D and Du, J and Ghosh, A and Love, B and Visotsky, E},
  journal={Nokia White Paper},
  year={2024}
}

@article{pala2023spectral,
  title={{Spectral-efficient RIS-aided RSMA URLLC: Toward mobile broadband reliable low latency communication (mBRLLC) system}},
  author={Pala, Sonia and Katwe, Mayur and Singh, Keshav and Clerckx, Bruno and Li, Chih-Peng},
  journal={IEEE Trans. Wireless Commun.	},
  volume={23},
  number={4},
  pages={3507--3524},
  year={2023},
  publisher={IEEE}
}

@article{khan2022urllc,
  title={{URLLC and eMBB in 5G industrial IoT: A survey}},
  author={Khan, Benish Sharfeen and Jangsher, Sobia and Ahmed, Ashfaq and Al-Dweik, Arafat},
  journal={IEEE Open J. Commun. Soc.	},
  volume={3},
  pages={1134--1163},
  year={2022},
  publisher={IEEE}
}

@article{bjornson2025enabling,
  title={{Enabling 6G performance in the upper mid-band by transitioning from massive to gigantic MIMO}},
  author={Bj{\"o}rnson, Emil and Kara, Ferdi and Kolomvakis, Nikolaos and Kosasih, Alva and Ramezani, Parisa and Salman, Murat Babek},
  journal={IEEE Open J. Commun. Soc.},
  year={2025},
  publisher={IEEE}
}

@inproceedings{tashman2025dynamic,
  title={{Dynamic Synergy: Leveraging RIS and Reinforcement Learning for Secure, Adaptive Underlay Cognitive Radio Networks}},
  author={Tashman, Deemah H and Cherkaoui, Soumaya},
  booktitle={2025 Global Information Infrastructure and Networking Symposium (GIIS)},
  pages={1--6},
  year={2025},
  organization={IEEE}
}

@article{tashman2026trustworthy,
  title={{Trustworthy AI-Driven Dynamic Hybrid RIS: Joint Optimization and Reward Poisoning-Resilient Control in Cognitive MISO Networks}},
  author={Tashman, Deemah H and Cherkaoui, Soumaya},
  journal={IEEE Trans. Netw. Serv. Manage.},
  year={2026},
  publisher={IEEE}
}

@article{rahim2026dual,
  title={{Dual-Tier IRS-Assisted Mid-Band 6G Mobile Networks: Robust Beamforming and User Association}},
  author={Rahim, Muddasir and Cherkaoui, Soumaya},
  journal={arXiv preprint arXiv:2602.00431},
  year={2026}
}

@article{rahim2026reliable,
  title={{Reliable IoT Communications in 6G Non-Terrestrial Networks with Dual RIS}},
  author={Rahim, Muddasir and Cherkaoui, Soumaya},
  journal={arXiv preprint arXiv:2602.00438},
  year={2026}
}

@inproceedings{cui20236g,
  title={6{G} Wireless Communications in 7–24 {GH}z Band: Opportunities, Techniques, and Challenges}, 
  author={Cui, Zhuangzhuang and Zhang, Peize and Pollin, Sofie},
  booktitle={2025 IEEE International Symposium on Dynamic Spectrum Access Networks (DySPAN)},
  pages={1--8},
  year={2025},
  organization={IEEE}
}

@article{you2022enabling,
  title={{Enabling smart reflection in integrated air-ground wireless network: IRS meets UAV}},
  author={You, Changsheng and Kang, Zhenyu and Zeng, Yong and Zhang, Rui},
  journal={IEEE Wireless Commun.	},
  volume={28},
  number={6},
  pages={138--144},
  year={2022},
  publisher={IEEE}
}

@inproceedings{heath2024beamsharing,
  title={{Beamsharing in mixed near-field/far-field MIMO systems for the upper mid-band}},
  author={Heath, Robert W and Gonz{\'a}lez-Prelcic, Nuria},
  booktitle={2024 IEEE 25th International Workshop on Signal Processing Advances in Wireless Communications (SPAWC)},
  pages={786--790},
  year={2024},
  organization={IEEE}
}

@article{tian2024mid,
  title={{Mid-Band Extra Large-Scale MIMO System: Channel Modeling and Performance Analysis}},
  author={Tian, Jiachen and Han, Yu and Li, Xiao and Jin, Shi and Wen, Chao-Kai},
  journal={IEEE Trans. Commun.	},
  year={2024},
  publisher={IEEE}
}

@article{kara2024reconfigurable,
  title={{Reconfigurable Intelligent Surfaces in Upper Mid-Band 6G Networks: Gain or Pain?}},
  author={Kara, Ferdi and Demir, {\"O}zlem Tu{\u{g}}fe and Bj{\"o}rnson, Emil},
  journal={arXiv preprint arXiv:2407.05754},
  year={2024}
}

@article{abdalla2020uavs,
  title={{UAVs with reconfigurable intelligent surfaces: Applications, challenges, and opportunities}},
  author={Abdalla, Aly Sabri and Rahman, Talha Faizur and Marojevic, Vuk},
  journal={arXiv preprint arXiv:2012.04775},
  year={2020}
}

@article{mohsan2023irs,
  title={{IRS-assisted UAV communications: A comprehensive review}},
  author={Mohsan, Syed Agha Hassnain and Li, Yanlong},
  journal={arXiv preprint arXiv:2306.15838},
  year={2023}
}

@article{farre2025dynamic,
  title={{Dynamic Beyond 5G and 6G Connectivity: Leveraging NTN and RIS Synergies for Optimized Coverage and Capacity in High-Density Environments}},
  author={Farr{\'e}, Valdemar and Estrada, Juan and Vega, David and Urquiza-Aguiar, Luis F and Peralvo, Juan A V{\'a}squez and Chatzinotas, Symeon},
  journal={arXiv preprint arXiv:2506.10900},
  year={2025}
}

@article{bui2025joint,
  title={{Joint Phase-Shift Design and Power Control for Near-and Far-Field Communications in Extremely Large RIS-aided UAV Networks}},
  author={Bui, Tinh T and Van Huynh, Dang and Nguyen, Long D and Jung, Haejoon and Duong, Trung Q},
  journal={IEEE Internet Things J.},
  year={2025},
  publisher={IEEE}
}

@article{rahim2025hierarchical,
  title={{Hierarchical IRS-assisted user association for sum-rate maximization in hybrid THz LEO-HAPS-terrestrial networks}},
  author={Rahim, Muddasir and Basit, Abdul and Selim, Bassant and Kaddoum, Georges},
  journal={IEEE Commun. Lett.},
  year={2025},
  publisher={IEEE}
}

@inproceedings{rahim2024user,
  title={{User association optimization for IRS-aided terahertz networks: A matching theory approach}},
  author={Rahim, Muddasir and Nguyen, Thanh Luan and Kaddoum, Georges and Do, Tri Nhu},
  booktitle={2024 IEEE Wireless Communications and Networking Conference (WCNC)},
  pages={1--6},
  year={2024},
  organization={IEEE}
}

@article{rahim2024jpusa,
  title={{JPUSA in coexistence of eMBB and URLLC services in multi-cell IRS-assisted Terahertz networks}},
  author={Rahim, Muddasir and Nguyen, Thanh Luan and Kaddoum, Georges},
  journal={IEEE Trans. Green Commun. Networking},
  year={2024},
  publisher={IEEE}
}

@article{rahim2024joint,
  title={{Joint power and user allocation in coexistence of eMBB and URLLC services}},
  author={Rahim, Muddasir and Nguyen, Thanh Luan and Do, Tri Nhu and Kaddoum, Georges},
  journal={IEEE Commun. Lett.},
  volume={28},
  number={9},
  pages={2186--2190},
  year={2024},
  publisher={IEEE}
}
\end{document}